\documentstyle[twoside]{article}

\catcode`\@=11
\long\def\@makefntext#1{ 
\protect\noindent \hbox to 3.2pt {\hskip-.9pt
$^{{\eightrm\@thefnmark}}$\hfil}#1\hfill} 

\def\thefootnote{\fnsymbol{footnote}}
 \def\@makefnmark{\hbox to 0pt{$^{\@thefnmark}$\hss}}  

\def\ps@myheadings{\let\@mkboth\@gobbletwo
\def\@oddhead{\hbox{} 
\rightmark\hfil\eightrm\thepage}
\def\@oddfoot{}\def\@evenhead{\eightrm\thepage\hfil 
\leftmark\hbox{}}\def\@evenfoot{}
\def\sectionmark##1{}\def\subsectionmark##1{}}

\oddsidemargin=\evensidemargin
\renewcommand{\thefootnote}{\fnsymbol{footnote}}

\newcounter{sectionc}\newcounter{subsectionc}\newcounter{subsubsectionc}
\renewcommand{\section}[1] {\vspace{12pt}\addtocounter{sectionc}{1}
\setcounter{subsectionc}{0}\setcounter{subsubsectionc}{0}\noindent
        {\tenbf\thesectionc. #1}\par\vspace{5pt}}
\renewcommand{\subsection}[1] {\vspace{12pt}\addtocounter{subsectionc}{1}
        \setcounter{subsubsectionc}{0}\noindent
        {\bf\thesectionc.\thesubsectionc. {\kern1pt \bfit #1}}\par\vspace{5pt}}
\renewcommand{\subsubsection}[1]
{\vspace{12pt}\addtocounter{subsubsectionc}{1}
        \noindent{\tenrm\thesectionc.\thesubsectionc.\thesubsubsectionc.
        {\kern1pt \tenit #1}}\par\vspace{5pt}}
\newcommand{\nonumsection}[1] {\vspace{12pt}\noindent{\tenbf #1}
        \par\vspace{5pt}}

\newcounter{appendixc}
\newcounter{subappendixc}[appendixc]
\newcounter{subsubappendixc}[subappendixc]
\renewcommand{\thesubappendixc}{\Alph{appendixc}.\arabic{subappendixc}}
\renewcommand{\thesubsubappendixc}
        {\Alph{appendixc}.\arabic{subappendixc}.\arabic{subsubappendixc}}

\renewcommand{\appendix}[1] {\vspace{12pt}
        \refstepcounter{appendixc}
        \setcounter{figure}{0}
        \setcounter{table}{0}
        \setcounter{lemma}{0}
        \setcounter{theorem}{0}
        \setcounter{corollary}{0}
        \setcounter{definition}{0}
        \setcounter{equation}{0}
        \renewcommand{\thefigure}{\Alph{appendixc}.\arabic{figure}}
        \renewcommand{\thetable}{\Alph{appendixc}.\arabic{table}}
        \renewcommand{\theappendixc}{\Alph{appendixc}}
        \renewcommand{\thelemma}{\Alph{appendixc}.\arabic{lemma}}
        \renewcommand{\thetheorem}{\Alph{appendixc}.\arabic{theorem}}
        \renewcommand{\thedefinition}{\Alph{appendixc}.\arabic{definition}}
        \renewcommand{\thecorollary}{\Alph{appendixc}.\arabic{corollary}}
        \renewcommand{\theequation}{\Alph{appendixc}.\arabic{equation}}
        \noindent{\tenbf Appendix \theappendixc #1}\par\vspace{5pt}}
\newcommand{\subappendix}[1] {\vspace{12pt}
        \refstepcounter{subappendixc}
        \noindent{\bf Appendix \thesubappendixc. {\kern1pt \bfit #1}}
        \par\vspace{5pt}}
\newcommand{\subsubappendix}[1] {\vspace{12pt}
        \refstepcounter{subsubappendixc}
        \noindent{\rm Appendix \thesubsubappendixc. {\kern1pt \tenit #1}}
        \par\vspace{5pt}}

\newcommand{\textlineskip}{\baselineskip=13pt}
\newcommand{\smalllineskip}{\baselineskip=10pt}

\def\eightcirc{
\begin{picture}(0,0)
\put(4.4,1.8){\circle{6.5}}
\end{picture}}
\def\eightcopyright{\eightcirc\kern2.7pt\hbox{\eightrm c}}

\def\abstracts#1#2#3{{
        \centering{\begin{minipage}{4.5in}\baselineskip=10pt\eightrm
        \centerline{ABSTRACT}
        \parindent=0pt #1\par
        \parindent=15pt #2\par
        \parindent=15pt #3
        \end{minipage} }\par}}

\newcommand{\bibit}{\nineit}

\renewenvironment{thebibliography}[1]                   
        {\ninerm
         \baselineskip=11pt                             
         \begin{list}{\arabic{enumi}.}
        {\usecounter{enumi}\setlength{\parsep}{0pt}
         \setlength{\leftmargin 17pt}{\rightmargin 0pt} 
         \setlength{\itemsep}{0pt} \settowidth          
        {\labelwidth}{#1.}\sloppy}}{\end{list}}

\newcounter{itemlistc}
\newcounter{romanlistc}
\newcounter{alphlistc}
\newcounter{arabiclistc}

\newcommand{\fcaption}[1]{
        \refstepcounter{figure}
        \setbox\@tempboxa = \hbox{\eightrm Fig.~\thefigure. #1}
        \ifdim \wd\@tempboxa > 5in
           {\begin{center}
        \parbox{5in}{\eightrm \smalllineskip Fig.~\thefigure. #1 }
            \end{center}}
        \else
             {\begin{center}
             {\eightrm Fig.~\thefigure. #1}
              \end{center}}
        \fi}

\newcommand{\tcaption}[1]{
        \refstepcounter{table}
        \setbox\@tempboxa = \hbox{\eightrm Table~\thetable. #1}
        \ifdim \wd\@tempboxa > 5in
           {\begin{center}
        \parbox{5in}{\eightrm\smalllineskip Table~\thetable. #1 }
            \end{center}}
        \else
             {\begin{center}
             {\eightrm Table~\thetable. #1}
              \end{center}}
        \fi}

\def\@citex[#1]#2{\if@filesw\immediate\write\@auxout    
        {\string\citation{#2}}\fi                       
\def\@citea{}\@cite{\@for\@citeb:=#2\do                 
        {\@citea\def\@citea{,}\@ifundefined             
        {b@\@citeb}{{\bf ?}\@warning
        {Citation `\@citeb' on page \thepage \space undefined}}
        {\csname b@\@citeb\endcsname}}}{#1}}

\newif\if@cghi
\def\cite{\@cghitrue\@ifnextchar [{\@tempswatrue
        \@citex}{\@tempswafalse\@citex[]}}
\def\citelow{\@cghifalse\@ifnextchar [{\@tempswatrue
        \@citex}{\@tempswafalse\@citex[]}}
\def\@cite#1#2{{$\null^{#1}$\if@tempswa\typeout
        {IJCGA warning: optional citation argument
        ignored: `#2'} \fi}}

\def\pmb#1{\setbox0=\hbox{#1}
        \kern-.025em\copy0\kern-\wd0
        \kern.05em\copy0\kern-\wd0
        \kern-.025em\raise.0433em\box0}

\def\fnt#1#2{\footnotetext{\kern-.3em
        {$^{\mbox{\scriptsize #1}}$}{#2}}}

\def\fpage#1{\begingroup
\voffset=.3in
\thispagestyle{empty}\begin{table}[b]\centerline{\footnotesize #1}
        \end{table}\endgroup}

\def\runninghead#1#2{\pagestyle{myheadings}
\markboth{{\eightit{\quad #1}}\hfill}{\hfill{\eightit{#2\quad}}}}

\font\tenbf=cmbx10
\font\tenit=cmti10
\font\tenrm=cmr10
\font\bfit=cmbxti10 at 10pt
 1
 1
 1

\font\ninerm=cmr9
\font\nineit=cmti9

\font\eightrm=cmr8
\font\eightit=cmti8

\def\qed{\hbox{${\vcenter{\vbox{                          
   \hrule height 0.4pt\hbox{\vrule width 0.4pt height 6pt
   \kern5pt\vrule width 0.4pt}\hrule height 0.4pt}}}$}}

\runninghead{The Role of Apparent Horizon on the Evolution of
$\ldots$} {The Role of Apparent Horizon on the Evolution of
$\ldots$}

\begin{document}
\normalsize\textlineskip
{\thispagestyle{empty}
\setcounter{page}{1}

\renewcommand{\thefootnote}{\fnsymbol{footnote}} 

{\noindent \footnotesize{MU-DM-App-94/16}}\\
{\footnotesize{gr-qc/9409024}}
\vspace*{0.88truein}

\fpage{1}
\centerline{\bf THE ROLE OF THE APPARENT HORIZON IN THE EVOLUTION OF}
\vspace*{0.035truein}
\centerline{\bf ROBINSON-TRAUTMAN EINSTEIN-MAXWELL SPACETIMES}
\vspace{0.37truein}
\centerline{\footnotesize ANTHONY W.\ C.\ LUN\ \ and\ \ ERNEST W.\ M.\ CHOW}
\vspace*{0.015truein}
\centerline{\footnotesize\it Department of Mathematics, Monash University,
Wellington Road}
\baselineskip=10pt
\centerline{\footnotesize\it Clayton, Victoria 3168,
Australia}
\vspace*{0.015truein}
\centerline{\footnotesize\it lun@vaxc.cc.monash.edu.au}
\centerline{\footnotesize\it ewmc@cosmo.maths.monash.edu.au}

\vspace{0.225truein}
\centerline{3rd September 1994}

\vspace*{0.21truein}
\abstracts{\noindent The `runaway solutions' of the Lorentz-Dirac
equation of a charged particle interacting with its own field in
classical electrodynamics are well-known. This type of self accelerated
phenomena also exists in the solutions of the Einstein-Maxwell equations
in general relativity. In particular, runaway solutions occur in a class
of simple models known as the `Asymptotically Flat Robinson-Trautman
Einstein-Maxwell' (AFRTEM) spacetimes. Consequently these spacetimes cannot
evolve to their unique regular steady state, viz. a charged non-rotating
black hole. This seems to contradict the established results that
charged non-rotating black holes are stable under first order perturbations.
We show that if an AFRTEM spacetime also possesses an apparent horizon,
then it has a Lyapunov functional.
This suggests that the evolution equations with additional constraints
arising from the apparent horizon would evolve stably to a charged
non-rotating black hole.}{}{}  %

\vspace*{-3pt}\textlineskip
\section{Introduction}
\noindent

The runaway solutions of the Lorentz-Dirac equation
of a charged particle interacting with its own field are considered
to be unphysical. This type of self accelerated motions has never been
observed down to the length scale of a classical  electron. Its non-existence
can be explained within the realm of classical electrodynamics.$^1$\, It is
not surprising that  runaway solutions of the Einstein-Maxwell
equations also exist in general relativity.$^2$  In particular,
runaway solutions of the evolution equations occur
in a class of simple models known as the {\em Asymptotically Flat
Robinson-Trautman Einstein-Maxwell\/} (AFRTEM) spacetimes.$^3$
Consequently this class of spacetimes cannot evolve to its unique
regular steady state, viz. a charged non-rotating black hole
represented by the Reissner-Nordstr\"om solution.
There are a number of reasons why such solutions are thought to be unphysical.
(1) It seems to contradict the result that first order perturbations
to a charged non-rotating black hole are stable.$^4$
(2) The corresponding charge-free models, which are the non-static
{\em Asymptotically Flat Robinson-Trautman\/} (AFRT) vacuum spacetimes,
emit gravitational waves then settle down to the unique steady state as a
non-rotating black hole represented by the Schwarzschild
solution.$^{5,6,7,8}$\,
(3) The runaway solutions suggest that the AFRTEM
spacetimes possess some kind of compact sources
which supply the release of an unlimited amount of
gravitational and electromagnetic radiation. (4) Numerical simulations
show that the Bondi-Sachs $4$-momentum
becomes infinite as the spacetimes evolve and this result can be
interpreted as self accelerated motion of the associated source.

We show that if we assume that the
AFRTEM spacetimes are dynamic black holes, then we can
eliminate the runaway solutions of the evolution equations.
Mathematically, the assumption gives rise to regularity conditions which
serve as inner boundary conditions `protecting' the past singularities.
In technical terms, we assume the existence of a past apparent
horizon in these spacetimes, which guarantees the presence of a particle
horizon. As a consequence, we are able to prove that these restricted
AFRTEM spacetimes possess a Lyapunov functional, which is the sum
of the square of the Bondi mass and the surface area of the apparent
horizon. This suggests that the AFRTEM evolution equations
with additional constraint equations arising from the apparent horizon
assumption would evolve stably to the Reissner-Nordstr\"om solution,
provided solutions to the nonlinear system of equations exist. We have
not proved the existence of solutions of this nonlinear system
of three functions of three independent variables, consisting of
three hypersurface equations, one hypersurface inequality and
one two-surface equation. However in the conclusion we indicate
that the linearised equations of these restricted spacetimes
are stable about the Reissner-Nordstr\"om solution (details in
preparation).

In Section 2 we briefly summarise the evolution equations
of the AFRTEM spacetimes and then show that
the linearisation of these equations about their steady-state solution
are unstable. We also consider a number of conserved quantities
arising from the evolution equations and prove that the
Bondi mass of the spacetime is a monotonically decreasing quantity.
For a more detailed discussion see Singleton.$^9$\, In Section 3 the
constraint equations arising from the apparent horizon assumption are
presented (see Chow and Lun $^{10}$ for derivation). In Section 4, we prove
that (1) the Bond mass of a restricted AFRTEM spacetime is positive,$^{10}$\,
(2) the surface area of the past apparent horizon
is monotonically decreasing, and (3) the sum of the square
of the Bondi mass and the surface area of the apparent
horizon is a Lyapunov functional, whose critical point is the
Ressiner Nordstr\"om solution.}{}{}

\textheight=7.8truein
\setcounter{footnote}{0}
\renewcommand{\thefootnote}{\alph{footnote}}

\section{The Asymptotically Flat RTEM Spacetimes}
\noindent
Physically the AFRTEM spacetimes can be thought of as
electrically charged non-rotating compact sources
emitting purely outgoing coupled gravitational and
electromagnetic radiations. A more general mathematical formulation
can be found in Kramer {\em et al.\/}.$^{11}$ In this paper we restrict
our discussion to the asymptotically flat subclass. The AFRTEM
spacetimes can be described by the metric

\begin{equation}
  ds^2 = \left( -2r(\ln P),_{u} +  K - {2M\over r}
         + {Q_{0}^{2}\over r^2}\right) du^2 +
                2dudr - {2r^2 d\zeta d{\bar{\zeta}} \over P^2 }
          \label{eq:metric}
\end{equation}

\noindent in
$(u,r,\zeta,{\bar\zeta}\,)$ coordinates. We use
``$\partial\,$'' or ``$,$'' to denote partial derivatives.
Here $M=M(u,\zeta,{\bar\zeta}\,)$
and $P=P(u,\zeta,{\bar\zeta}\,)$ are real-valued functions. $Q_{0}$ is a real
constant corresponding to the electric charge of the spacetime.
$K:=\Delta(\ln P)\,$, where
$\Delta:=2P^2 \partial_{\zeta \bar\zeta}\,$ is the Laplacian on the
closed two-surface $S_{u,r}$ which has the topology of a two-sphere $S^2$.
Geometrically $K$ is the Gaussian curvature of $S_{u,r}$.
The spacetime is foliated by these two-surfaces as $u$ and $r$ vary.
$u$ is the retarded time and $u=u_0=\hbox{constant}$ is a future null
cone whose generators are a congruence of diverging, shearfree null
geodesics affinely parametrised by $r$.
The variables $M$ and $P$ can be interpreted as the mass-energy of the
spacetime. Information about the rest mass and electromagnetic radiations
of the spacetime is prescibed by $M$ while information about
gravitational radiations is given by $P\/$.$^{10}$
After partial integration of the Einstein-Maxwell equations, the
residual field equations can be written as :

\begin{eqnarray}
     (\ln P),_u & = & -{1\over {4Q_{0}^2}} \Delta M   \label{eq:rt1}  \\
           M,_u & = & -{3M\over {4Q_{0}^2}} \Delta M + {1\over 4}\Delta K -
           {P^2\over 2Q_{0}^2} M,_{\zeta} M,_{\bar\zeta}   \label{eq:rt2}
\end{eqnarray}

\noindent Since the evolution equations
(\ref{eq:rt1}) and (\ref{eq:rt2})
are independent of $r$, the whole spacetime is determined by the
evolution of $M$ and $P$ with respect to the retarded time $u$
on a $r=r_{0}=\hbox{constant}$ hypersurface. Without any loss of
generality we take $r_{0}=1$ in the following.

\subsection{Linearised Evolution equations}
\noindent
In contrast to the AFRT vacuum spacetimes, the AFRTEM evolution
equations (\ref{eq:rt1}) and (\ref{eq:rt2}) are linearly unstable about the
Reissner Nordstr\"om solution, which is the unique regular steady state
solution of these equations. The steady state solution is given by

\begin{equation}
        M=M_{0}=\hbox{constant,\ \ \ }
        P=P_{0}:={1\over \sqrt{2}\/}(a + \bar{b} \zeta + b \bar{\zeta} +
                 c \zeta \bar{\zeta}\/)               \label{eq:rn}
\end{equation}

\noindent where
$ac-b\bar{b}=1$ and $M_{0}$ is the mass of the Reissner-Nordstr\"om
solution. In terms of $P_{0}$, the metric of the two-sphere $S^{2}$ is
$ g_{0} = 2P_{0}^{-2} d\zeta d\bar{\zeta}\/ $. The condition
$ac-b\bar{b}=1$ normalises the constant Gaussian curvature of
$S^2$ to $K=K_{0}=1$.

Following Singleton,$^9$ we linearise (\ref{eq:rt1}) and (\ref{eq:rt2})
about the steady state solution (\ref{eq:rn}).
Let $P=P_{0}(1+\epsilon\tilde{G}\/)$ and $M=M_{0}(1+\epsilon\tilde{E}\/)$,
where the parameter $\epsilon$ is a first order quantity.
Physically $\tilde{G}$ and $\tilde{E}$
describe the purely outgoing gravitational and electromagnetic
radiations, respectively.
The linearised equations are

\begin{eqnarray}
\tilde{G},_{u} &=& -{M_{0}\over {4Q_{0}^2}}
                           \Delta_{0} \tilde{E}      \label{eq:lrt1}   \\
\tilde{E},_{u} &=& -{3M_{0}\over {4Q_{0}^2}} \Delta_{0} \tilde{E} +
         {1\over {4M_{0}}}(\Delta_{0}+2)\Delta_{0} \tilde{G}  \label{eq:lrt2}
\end{eqnarray}

\noindent where
$\Delta_{0}=2P_{0}^2 \partial_{\zeta \bar\zeta}\/$ denotes the Laplacian
on $S^2$.
Decomposition into spherical harmonics $\tilde{G} = \Sigma
G_{\ell,m}Y_{\ell,m}$ and $\tilde{E} = \Sigma E_{\ell,m}Y_{\ell,m}$ reduces
the linearised  equations (\ref{eq:lrt1}) and (\ref{eq:lrt2}) to a
linear autonomous system in $G_{\ell,m}$ and $E_{\ell,m}\/$:

\begin{displaymath}
    {d \over {du}}
                  \left(\begin{array}{c}
                   G_{\ell,m} \\  E_{\ell,m}  \end{array}  \right)
    =
    \left(\begin{array}{cc}
    0 & {M_{0}\over {4Q_{0}^2}}\ell(\ell+1)  \\
        {1\over{4M_{0}}}(\ell+2)(\ell+1)\ell(\ell-1) &
        {3M_{0}\over{4Q_{0}^2}} \ell(\ell+1) \end{array}
    \right)
    \left(\begin{array}{c}
     G_{\ell,m}  \\  E_{\ell,m}     \end{array}  \right)
\end{displaymath}

\noindent
The eigenvalues of
the linear system are given by

\begin{equation}
\lambda_{\ell,\pm} = {3M_{0}\over{8Q_{0}^2}}\ell(\ell+1) \left(1\pm
\sqrt{1+{4Q_{0}^2 (\ell+2)(\ell-1)\over {9 M_{0}^2}}} \right)  \label{eq:eigen}
\end{equation}

\noindent Thus all modes
with $\ell\geq 1$ of the linearised equations are unstable.
The critical point of the system is a saddle.
Therefore the nonlinear equations will also be unstable about
the steady state solution.$^{12}$\,
This would pose a rather serious problem for a physically meaningful
interpretation of the AFRTEM spacetimes. Furthermore it apparently
contradicts the established results affirming the stability
of the Reissner-Nordstr\"om black hole against linear perturbations.$^4$
We note, however, that for any mode $\ell\/$, there is a stable submanifold
given by a linear combination of the eigenvectors corresponding to
$\lambda_{\ell,-}\/$.  For $\ell=1\/$ (i.e. the dipole modes), the eigenvalue
$\lambda_{1,+} = {3M_{0}\over{2Q_{0}^2}}$ is the same as the exponent
in the runaway solution of the radiation reaction equation of motion for
a free particle in classical electrodynamics.$^2$\, Hence this justifies
calling these unstable solutions of the AFRTEM
evolution equations runaway solutions of the Einstein-Maxwell equations.

\subsection{Conserved Quantities}
\noindent
{}From the evolution equations (\ref{eq:rt1}) and (\ref{eq:rt2}),
one can derive a number of conserved quantities. It is more convenient to
prove these and subsequent results if we `factor out' the
two-sphere geometry from the evolution equations by defining
$e^{-\Lambda}:= P/P_{0}$. The metric of the two-surface $S_{u,1}$
is conformal to the metric of $S^{2}$ with the conformal factor
$e^{2\Lambda}\/$ so that

\begin{displaymath}
         g_{1} = {2e^{2\Lambda}d\zeta d\bar\zeta \over P_{0}^2}
               = e^{2\Lambda}g_{0}
\end{displaymath}

\noindent It is important to note that, unlike $P$ which is singular at
a point (at least one) on $S_{u,1}$ due to the non-trivial topology of
$S^2$, the conformal factor $e^{\Lambda}$ is regular
(cf. the form of $P_{0}$ in
Eq. (\ref{eq:rn})\,). The surface elements of $S^{2}$
and $S_{u,1}$ are given, respectively, by

\begin{displaymath}
\omega_{0}={{i d\bar\zeta \wedge d\zeta}\over {P_{0}^{2}}}\,,\;\;\;\;
\omega_{1}=e^{2\Lambda}\omega_{0}
\end{displaymath}

\noindent
The evolution equations (\ref{eq:rt1}) and (\ref{eq:rt2})
are now expressed in terms of regular functions $\Lambda$ and
$M$ on $S^{2}$ as :

\begin{eqnarray}
e^{2\Lambda} \Lambda,_u &=& {1\over{4Q_{0}^2}} \Delta_{0} M  \label{eq:rt1'} \\
     e^{2\Lambda}  M,_u &=& -{3M\over {4Q_{0}^2}} \Delta_{0} M
                    + {1\over 4}\Delta_{0} K - {{(P_{0}
M,_{\zeta})(P_{0} M,_{\bar{\zeta}})} \over {2Q_{0}^{2}}} \label{eq:rt2'}
\end{eqnarray}

\noindent
We define three quantities on the closed two-surface $S_{u,1}\/$,
its surface area ${\cal A}_{S_{u,1}}$, its Euler characteristic $\chi\,$ and
its `irreducible mass' ${\cal M}$ as follows :

\begin{eqnarray}
        {\cal A}_{S_{u,1}} &:=& \int_{S_{u,1}} \omega_{1}
                  = \int_{S^2} e^{2\Lambda} \omega_{0}   \label{eq:area1} \\
                      \chi &:=& {1\over {2\pi}} \int_{S_{u,1}} K \omega_{1}
={1\over {2\pi}} \int_{S^2} K e^{2\Lambda} \omega_{0}    \label{eq:euler} \\
                  {\cal M} &:=& \int_{S_{u,1}} M \omega_{1}
             = \int_{S^2} M e^{2\Lambda} \omega_{0}      \label{eq:mass1}
\end{eqnarray}

\noindent
The conservation of the surface area ${\cal A}_{S_{u,1}}\/$
(i.e. ${d\over du}{\cal A}_{S_{u,1}}\,=0\,$) follows from
(\ref{eq:rt1'}) by integrating it over $S^2$ and applying
Stokes' theorem to the right side.
In terms of $\Lambda$, the Gaussian curvature

\begin{equation}
K=e^{-2\Lambda\/}(1-\Delta_{0} \Lambda\/)
 =e^{-2\Lambda\/}+e^{-\Lambda}\Delta_{0} e^{-\Lambda}-
  2(P_{0} e^{-\Lambda\/}),_{\zeta}(P_{0} e^{-\Lambda\/}),_{\bar\zeta}
                        \label{eq:gc}
\end{equation}

\noindent Substituting
(\ref{eq:gc}) into (\ref{eq:euler}) gives the topological invariant

\begin{equation}
\chi ={1\over {2\pi}} \int_{S^2} (1-\Delta_{0} \Lambda\,)\,\omega_{0}
     ={1\over {2\pi}} \int_{S^2} \omega_{0} = 2          \label{eq:euler2}
\end{equation}

\noindent This implies the conservation of the Euler characteristic $\chi$
and that $S_{u,1}$ is
a topological $S^2$ for all retarded time $u$. Alternately,
from (\ref{eq:rt1'}) and the definition of $K$, one can
show that

\begin{eqnarray*}
            K,_{u} &=& -{1 \over {4Q_{0}^2}}(\Delta + 2K)\Delta M   \\
(e^{2\Lambda}K),_u &=& -{1 \over {4Q_{0}^2}}
                      \Delta_{0}(e^{-2\Lambda}\Delta_{0} M)
\end{eqnarray*}

\noindent ${d\over{du}}\chi=0$ follows from the latter equation.
Hence the result derived from the evolution equations is consistent
with that from (\ref{eq:euler2}).
Since $\chi$ is an integer, a non-constant $\chi$ would indicate
pathological and unphysical behaviour of the field equations.
Combining (\ref{eq:rt1'}) and (\ref{eq:rt2'}) gives

\begin{displaymath}
     (e^{2\Lambda}M),_{u}={1 \over 4}\Delta_{0}(K-{M^{2}\over{2Q_{0}^2}})
\end{displaymath}

\noindent which implies the conservation of the irreducible mass ${\cal M}$.
The proof is similar to that of (\ref{eq:mass1}).

\subsection{Bondi Mass}
\noindent
In an asymptotically flat spacetime, the notion of the
energy-momentum at a given retarded time exists, viz. the
Bondi-Sachs $4$-momentum. One can consider the spacetime geometry
as one approaches null infinity ${\cal I}^+$ on an asymptotically
null hypersurface. This would enable one to discuss quantitatively
the amount of energy and momentum carried away by coupled
gravitational and electromagnetic radiation.
Using Penrose's conformal technique (see
Penrose and Rindler $^{13}$\,), the Bondi-Sachs
$4$-momentum of the AFRTEM spacetime can be shown to be $^{9,10}$

\begin{equation}
{\cal P}_{a} = {1\over 4\pi} \int_{S_{u,\infty}}
                             M e^{\Lambda}\xi_{a}\, \omega_{1}
={1\over 4\pi} \int_{S^2} M e^{3\Lambda}\xi_{a}\, \omega_{0}  \label{eq:4mom}
\end{equation}

\noindent where

\begin{displaymath}
\xi_{a}=(1,x,y,z\/),\;\; x={{\zeta + \bar\zeta}\over{\zeta{\bar\zeta}+1}},\;\;
      y=-{{i(\zeta - \bar\zeta)}\over{\zeta{\bar\zeta}+1}},\;\;
      z={{\zeta{\bar\zeta}-1}\over{\zeta{\bar\zeta}+1}}\;\;
\end{displaymath}

\noindent corresponds to the asymmptotic symmetries generating the
translations,
and $S_{u,\infty}$ is a two-sphere $S^2$ at ${\cal I}^{+}$ on a
$u=\hbox{constant}$ future null cone (called a `cut' of ${\cal I}^+\/$).
The time (first) component of ${\cal P}_{a}$ is the Bondi mass which
is denoted by ${\cal M}_{B}$, i.e.

\begin{equation}
{\cal P}_{t}:={\cal M}_{B} = {1\over 4\pi} \int_{S_{u,\infty}}
                             M e^{\Lambda}\, \omega_{1}
={1\over 4\pi} \int_{S^2} M e^{3\Lambda}\, \omega_{0}  \label{eq:bm}
\end{equation}

\noindent The components of the Bondi-Sachs $4$-momentum are
written as ${\cal M}_{B}\,$, ${\cal P}_{x}\,$, ${\cal P}_{y}$ and
${\cal P}_{z}\,$.
The Bondi mass is not a conserved quantity, indeed,
it is monotonically decreasing.

\vspace*{12pt}
\noindent
{\tenbf Proposition 1:} The Bondi mass of an AFRTEM spacetime is a
monotonically decreasing function of the retarded time $u\,$.

\vspace*{12pt}
\noindent
{\tenbf Proof:} Firstly, we use (\ref{eq:gc}) and the identity

\begin{displaymath}
[\Delta_{0} (e^{-\Lambda})],_\zeta =
[2P_{0}^2 (e^{-\Lambda}),_{\zeta \bar\zeta}],_{\zeta} =
2\left[ P_{0}^2\,\partial_{\bar\zeta}\!\left({1 \over P_{0}^{2}}
[P_{0}^2 (e^{-\Lambda}),_{\zeta}],_{\zeta} \right)
- (e^{-\Lambda}),_{\zeta} \right]
\end{displaymath}

\noindent
to expand $\Delta_{0} K$ in terms of $e^{-\Lambda\/}$ :

\begin{equation}
\Delta_{0} K = 4 \left\{e^{-\Lambda}P_{0}^2
\left[P_{0}^2\,\partial_{\bar\zeta}\!\left(
{1 \over P_{0}^{2}}[P_{0}^2 (e^{-\Lambda}),_{\zeta}],_{\zeta} \right)
\right] - [P_{0}^2 (e^{-\Lambda}),_{\zeta}],_{\zeta}
[P_{0}^2 (e^{-\Lambda}),_{\bar\zeta}],_{\bar\zeta}\right\}
               \label{eq:lgc}
\end{equation}

\noindent Then (\ref{eq:rt1'}) and (\ref{eq:rt2'}) imply

\begin{eqnarray}
(e^{3\Lambda}  M),_{u} = e^{\Lambda}\left({1\over 4}\Delta_{0} K - {{(P_{0} M,
_{\zeta})(P_{0} M,_{\bar{\zeta}})} \over {2Q_{0}^{2}}}\right)  \label{eq:rt2''}
\end{eqnarray}

\noindent
Substitute (\ref{eq:lgc}) into (\ref{eq:rt2''}), then integrate over
$S^2$ and apply Stokes' theorem to the right side. We obtain

\begin{equation}
 {d\over du}{\cal M}_B = -{1\over 4\pi} \int_{S^2} e^{\Lambda}\left(
           [P_{0}^2\/(e^{-\Lambda\/}),_{\zeta\/}],_{\zeta\/}
           [P_{0}^2\/(e^{-\Lambda\/}),_{\bar\zeta\/}],_{\bar\zeta\/} +
  {{(P_{0} M,_{\zeta})(P_{0} M,_{\bar{\zeta}})} \over {2Q_{0}^{2}}}\right)\/
  \omega_{0}  \leq 0                         \label{eq:dbm}
\end{equation}

\noindent since the integrand is non-negative.
\qed\,.

\vspace*{12pt}
The result is a special case of the positive mass loss
theorem due to Bondi et al.$^{14}$ and Sachs.$^{15}$\,
Using H\"older's inequality one can show that

\begin{equation}
\left(\int_{S^2} \omega_{0}\/\right)^{1\over 3\/}
\left(\int_{S^2} e^{3\Lambda} \omega_{0}\/\right)^{2\over 3\/} \geq
\int_{S^2} e^{2\Lambda} \omega_{0}        \label{eq:pbm}
\end{equation}

\noindent
In an AFRT vacuum spacetime, where $M=M_{0}$ is a positive constant,
the conservation of surface area (\ref{eq:pbm}) implies that
the Bondi mass of an AFRT vacuum spacetime is positive :
${\cal M}_{B} \geq M_{0}\/$.$^9$\, However unlike the vacuum case,
the expression (\ref{eq:bm}) for the Bondi mass of an AFRTEM spacetime is
not manifestly positive, since $M$ is no longer a positive constant.
In view of the positive Bondi mass theorem of Gibbons
et al.,$^{16}$ which assumes the presence of an apparent horizon
when spacetime singularities exist, such as the
$r=0$ singularity in the AFRT and the AFRTEM spacetimes,
hence the positivity of ${\cal M}_{B}$ arising from
(\ref{eq:pbm}) is a stronger result than the positive mass
theorem. This is due to the special features that in an AFRT vacuum
spacetime $M$ is constant and the inequality (\ref{eq:pbm}) holds.
In the next section, we will investigate how the
presence of an apparent horizon in a AFRTEM spacetime can lead
to the positivity of the associated Bondi mass.{}{}

\section{Apparent Horizon}
\noindent
In this section we assume that in an AFRTEM spacetime a past apparent
horizon exists and we will derive the governing equations.
We define an outer apparent horizon
${\cal H}$ to be a non-timelike hypersurface $r=\Re(u,\zeta,\bar\zeta\/)$
such that, on each $u=u_0$ null hypersurface, the spacelike two-surface
$r=\Re(u_0,\zeta,\bar\zeta\/)$ is a past marginally trapped suface
denoted by ${\cal T}_{u_0}$. These closed two-surfaces
${\cal T}_{u}$ can be identified by the
local conditions (1) that the ingoing future directed congruence of
orthogonal null geodesics must have vanishing divergence, and (2) that
the outgoing future directed congruence of orthogonal null
geodesics is diverging.$^{17}$\/ Hence the closed two-surfaces ${\cal T}_u$
foliate the hypersurface ${\cal H}$\,.
By the singularity theorems of Penrose $^{18}$ and Hawking $^{19}$,
the existence of an outer past apparent horizon guarantees
a `white hole'\/. By a `white hole' we mean
a past spacetime singularity `hidden' behind
a null hypersurface called the `particle horizon'\/, which is defined
as the boundary of the region to which particles or photons from the
past infinity can reach.

\subsection{Newman-Penrose Quantities}
\noindent In an AFRTEM spactime,
we can choose a null tetrad $(l^a, n^a, m^a, {\bar m}^a\/)$
such that $l^a$ is tangent to a congruence of
outgoing future directed null geodesics which is diverging, shearfree,
twistfree and affinely parametrised by $r\/$.
The contravariant form of the metric (\ref{eq:metric}) is given by
$g^{ab}:=l^a n^b + n^a l^b - m^a {\bar m}^b - {\bar m}^a m^b$ with

\begin{eqnarray}
l^a &=& \partial_r    \nonumber    \\
n^a &=& \partial_u - \left(H - {P^2 \Re,_{\zeta} \Re,_{\bar \zeta} \over
r^2}\right)\partial_r +
{P^2 \Re,_{\bar \zeta} \over r^2}\partial_\zeta +
{P^2 \Re,_\zeta \over r^2}\partial_{\bar \zeta}  \nonumber    \\
m^a &=&  {P\over r}(\partial_{\bar \zeta} +  \Re,_{\bar \zeta}
\partial_r)    \nonumber     \\
{\overline m}^a &=&  {P\over r}(\partial_\zeta +  \Re,_\zeta
\partial_r)       \label{eq:tetrad}
\end{eqnarray}

\noindent
Here $H= -r(\ln P),_u + {1\over 2}K -{M\over r}+{Q_{0}^2\over {2r^2}}\/$
and $\Re=\Re(u,\zeta,\bar\zeta)$ is a real-valued function whose
presence dose not affect the metric.
The nonvanishing Newman-Penrose $^{20}$ quantities are :

\begin{eqnarray}
\rho &=& -{1\over r}\/,\;\;\;
\tau = {\overline \pi} = {\overline \alpha}+\beta
= -{P\Re,_{\bar \zeta} \over r^2}\/,\;\;\;
\beta = -{P,_{\bar \zeta} \over 2r}\/,   \nonumber     \\
\lambda &=& (\partial_\zeta +  \Re,_\zeta \partial_r)
\left({P^2 \Re,_\zeta \over r^2}\right)\/,   \nonumber   \\
\mu &=&  -{1 \over 2r} \left[ K
 - {2 \over r}(M + P^2 \Re,_{\zeta {\bar \zeta}})
+{1 \over r^2}({Q_{0}^{2}}+{2P^2 \Re,_\zeta \Re,_{\bar \zeta}})\right]\/,
\nonumber       \\
\gamma &=&  -{1 \over 2} \left[ (\ln P),_{u},-
{1 \over r^2}(M + {P P,_\zeta \Re,_{\bar \zeta}}
                   - {P P,_{\bar \zeta} \Re,_\zeta})
+ {1 \over r^3}({Q_{0}^{2}}+{2P^2 \Re,_\zeta \Re,_{\bar \zeta}}) \right]\/,
\nonumber       \\
\nu &=& {P \over r}(\partial_\zeta +  \Re,_\zeta \partial_r)
\left(\Re,_u -r(\ln P),_u + {K \over 2} -{M\over r}+{Q_{0}^2\over {2r^2}}
+{P^2 \Re,_{\zeta} \Re,_{\bar \zeta} \over r^2}\right)\/,  \nonumber     \\
%
\phi_1 &=& {Q_0 \over 2r^2}\/,\;\;\;
\phi_2 = {P \over 2Q_{0}r}
         \left(M,_\zeta + {2Q_{0}^2 \Re,_\zeta \over r^2}\right)\/,\;\;\;
\Psi_2 = -{1 \over r^3}\left(M - {Q_{0}^2 \over r}\right)\/,    \nonumber    \\
\Psi_3 &=& -{P K,_\zeta \over 2r^2}+{3 P M,_\zeta \over 2r^3}
          -{3 P M \Re,_\zeta \over r^4}
          +{3 Q_{0}^2 P \Re,_\zeta \over r^5}\/,   \nonumber   \\
\Psi_4 &=& -{[P^2 (\ln P),_{u \zeta}],_{\zeta} \over r}
         +{(P^2 K,_\zeta),_\zeta \over 2r^2}
         -{[(P^2 M,_\zeta),_\zeta + 2P^2 K,_\zeta \Re,_\zeta] \over r^3}
  \nonumber         \\
       & &  +{6 P^2 M,_\zeta \Re,_\zeta \over r^4}
         -{6 P^2 M (\Re,_\zeta)^2 \over r^5}
         +{6 Q_{0}^2 P^2 (\Re,_\zeta)^2 \over r^6}   \label{eq:spcf}
\end{eqnarray}

\noindent Here we have chosen $\phi_{ab}=2\phi_{a}{\overline{\phi}_b}$
in the Newman-Penrose equations.

\subsection{Apparent Horizon Equations}
\noindent
Consider a hypersurface ${\cal H}$ given by $r=\Re(u,\zeta,\bar\zeta)$.
We use ${\widetilde{\cal Q}}$ to denote the restriction of the
quantity ${\cal Q}$ on the hypersurface ${\cal H}\/$.
Let ${\widetilde N}_{a}$
be a one-form on ${\cal H}$ defined by
${\widetilde N}_{a} = -\Re,_u du + dr - \Re,_{\zeta} d\zeta - \Re,_{\bar\zeta}
d{\bar\zeta}\/$. Hence ${\widetilde N}^{a}:=g^{ab}{\widetilde N}_{b}$ is a
vector orthogonal to ${\cal H}$ and is given by

\begin{eqnarray}
{\widetilde N}^a &=& {\widetilde n}^a  - \left(\Re_{u} + {\widetilde{H}} +
{P^2 \Re,_{\zeta} \Re,_{\bar \zeta} \over \Re^2}\right){\widetilde l}^a
                 \nonumber        \\
              &=& \partial_u - (\Re_{u} + 2{\widetilde{H}})\partial_r +
{P^2 \Re,_{\bar \zeta} \over \Re^2}\partial_\zeta +
{P^2 \Re,_\zeta \over \Re^2}\partial_{\bar \zeta}
                 \label{eq:normal}
\end{eqnarray}

\noindent From (\ref{eq:tetrad}), the complex null vectors
${\widetilde m}^a$ and ${\widetilde{\overline m}}^a$ are tangent to
the two-sufaces ${\cal T}_{u}$ which foliate the hypersurface ${\cal H}$.
According to (\ref{eq:tetrad}), the null vectors ${\widetilde l}^a$ and
${\widetilde n}^a$ are orthogonal to ${\cal T}_{u}$. It is
straightforward to check that the vector

\begin{eqnarray}
{\widetilde Z}^a &=& {\widetilde n}^a  + \left(\Re_{u} + {\widetilde{H}} +
{P^2 \Re,_{\zeta} \Re,_{\bar \zeta} \over \Re^2}\right){\widetilde l}^a
                 \nonumber                           \\
&=&\partial_u + \left(\Re_{u}+{2P^2 \Re,_\zeta \Re,_{\bar \zeta} \over \Re^2}
\right)\partial_r + {P^2 \Re,_{\bar \zeta} \over \Re^2}\partial_\zeta +
{P^2 \Re,_\zeta \over \Re^2}\partial_{\bar \zeta}
                 \label{eq:triad}
\end{eqnarray}

\noindent is orthogonal to ${\widetilde N}^a$ and therefore is tangent
to the hypersurface ${\cal H}$.
{}From (\ref{eq:normal}) and (\ref{eq:triad}), one obtains the
`magnitude' of ${\widetilde N}^a$ and of ${\widetilde Z}^a$

\begin{equation}
{\widetilde N}_a{\widetilde N}^a=-{\widetilde Z}_a{\widetilde Z}^a
=-2\left(\Re_{u} + {\widetilde{H}} +
{P^2 \Re,_{\zeta} \Re,_{\bar \zeta} \over {\Re}^2}\right)  \label{eq:length}
\end{equation}

\noindent It is more convenient to work with non-null normalised vectors.
Thus when $\Re_{u} + {\widetilde{H}} +
(P^2 \Re,_{\zeta} \Re,_{\bar \zeta})/{\Re}^2 \neq 0$, we use

\begin{displaymath}
{\widetilde k}^a={{\widetilde N}^a \over \sqrt{|{\widetilde N}_b
                  {\widetilde N}^b|}}\,,\;\;\;
{\widetilde z}^a={{\widetilde Z}^a \over \sqrt{|{\widetilde Z}_b
                  {\widetilde Z}^b|}}
\end{displaymath}

\noindent In this case the spacelike hypersurface ${\cal H}$ is
spanned by $[{\widetilde z}^a,{\widetilde m}^a,{\widetilde{\overline m}}^a]$
with unit normal vector ${\widetilde k}^a$.

(\ref{eq:tetrad}) and (\ref{eq:spcf}) imply that the null
vector ${\widetilde l}^a$ is
future directed, outward pointing, geodetic (${\widetilde{\kappa}}=0$),
diverging (${\widetilde{\rho}}=-{1 \over \Re}$), null and orthogonal to the
two-surface ${\cal T}_u$. The conditions for
${\cal H}$ to be an outer past apparent
horizon in an AFRTEM spacetime reduce to (1) the null vector
${\widetilde n}^a$ has vanishing divergence, i.e.
${\widetilde{\mu}}+{\widetilde{\overline \mu}}=0$, and
(2) the apparent horizon ${\cal H}$ is non-timelike, $^{17}$ i.e.
the normal vector ${\widetilde N}^a$ is causal and hence
${\widetilde N}_a{\widetilde N}^a \geq 0$.
{}From (\ref{eq:spcf}) and (\ref{eq:length}), these conditions become,
respectively,

\begin{eqnarray}
K - {2M \over \Re} + {Q_{0}^2 \over \Re^2} - \Delta(\ln \Re) &=& 0
       \label{eq:mu}   \\
2\Re\,\partial_{u}\!\left(\ln {\Re \over P}\right) +
     K - {2M \over \Re} + {Q_{0}^2 \over \Re^2} +
     {2P^2 \Re,_\zeta \Re,_{\bar \zeta} \over \Re^2}
& \leq & 0                        \label{eq:inequality}
\end{eqnarray}

Since $\mu$ is real, (\ref{eq:mu}) implies
${\widetilde{\mu}}=0$. Consequently the directional derivative
of ${\widetilde{\mu}}$ along the non-timelike
vector ${\widetilde Z}^a$ tangent to ${\cal H}$ must vanish.
(\ref{eq:triad}) and (\ref{eq:length}) then imply

\begin{equation}
{\widetilde Z}^a \nabla_a \widetilde{\mu} =
{\widetilde n}^a \nabla_a \widetilde{\mu} -
{1 \over 2}({\widetilde N}_b{\widetilde N}^b\,)
{\widetilde l}^a \nabla_a \widetilde{\mu} = 0      \label{eq:dd}
\end{equation}

\noindent Substituting (\ref{eq:spcf}) into
the Newman-Penrose equations $(4.11.12.a')$ and $(4.11.12f')$
in Penrose and Rindler $^{20}$ gives, respectively,

\begin{eqnarray}
{\widetilde n}^a \nabla_a \widetilde{\mu}&=&
{\widetilde m}^a \nabla_a \widetilde{\nu} +
\widetilde{\nu}(-\widetilde{\tau} + \widetilde{\overline \alpha} +
3\widetilde{\beta}) + \widetilde{\pi} \widetilde{\overline \nu} -
\widetilde{\lambda} \widetilde{\overline \lambda} -
2\widetilde{\phi}_2 \widetilde{\overline \phi}_2    \nonumber      \\
&=& -\partial_{\zeta}\!\left({P^2 \Re,_\zeta \over \Re^2}\right)
    \partial_{\bar \zeta}\!\left({P^2 \Re,_{\bar\zeta} \over \Re^2}\right) -
{P^2 \over 2Q_{0}^2 \Re^2}\partial_{\zeta}\!\left(M-{2Q_{0}^2 \over \Re}\right)
\partial_{\bar \zeta}\!\left(M-{2Q_{0}^2 \over \Re}\right)  \nonumber   \\
& & -{P^2 \over 2\Re}\left[\,\partial_{\zeta {\bar \zeta}}\!\left(
    {{\widetilde N}_a{\widetilde N}^a \over \Re}\right) -
    ({\widetilde N}_a{\widetilde N}^a)
    \,\partial_{\zeta {\bar \zeta}}\!\left({1 \over \Re}\right)
    \right]                     \label{eq:np1}        \\
{\widetilde l}^a \nabla_a \widetilde{\mu}&=&
{\widetilde m}^a \nabla_a \widetilde{\pi} +
\widetilde{\pi}(\widetilde{\overline \pi} - \widetilde{\overline \alpha} +
\widetilde{\beta}) + \widetilde{\Psi}_2    \nonumber     \\
&=&{P^2 \over \Re} \partial_{\zeta{\bar \zeta}}\left({1 \over \Re}\right)
- {M\over \Re^3} + {Q_{0}^2 \over \Re^4}             \label{eq:np2}
\end{eqnarray}

\noindent In (\ref{eq:np1}) we have used $\widetilde{\nu}=-{P \over 2\Re}
({\widetilde N}_a{\widetilde N}^a),_\zeta\,$.
Combine (\ref{eq:dd}), (\ref{eq:np1}) and (\ref{eq:np2})
to give

\begin{equation}
{1\over \Re}\Delta\left({{\widetilde N}_a{\widetilde N}^a \over \Re}\right)
-2 \widetilde{\Psi}_2({\widetilde N}_a{\widetilde N}^a)
=  4(\widetilde{\lambda} \widetilde{\bar\lambda} + 2\widetilde{\phi}_2
\widetilde{\overline {\phi}}_2)                     \label{eq:h3}
\end{equation}

\noindent
An AFRTEM spacetime with apparent horizon can be described then by the
evolution equations (\ref{eq:rt1'}) and (\ref{eq:rt2'}) together with
the constraint equations (\ref{eq:mu}) and (\ref{eq:h3}), and the
constraint inequality (\ref{eq:inequality}).
%
%
%
%
%
%
%
%
%
%
%

\section{Lyapunov Functional}
\noindent
In this section we will show that the existence of a past apparent horizon
in an AFRTEM spacetime implies that the Bondi Mass of the spacetime will
always be positive.  This can be regarded as a special case of the
general Bondi Mass positivity result for charged black holes.$^{16,21}$\,
In fact, all that is required for this result to hold on any given slice
$u=u_0$
in an AFRTEM spacetime is that there be a marginally trapped surface on that
slice; i.e. only (\ref{eq:mu}) is used. The inequality (\ref{eq:inequality})
guarantees that the area of the apparent horizon monotonically
decreases with increasing $u$.  These results will be combined to show
the existence of a Lyapunov functional for the spacetime. We also use
$e^{\Phi}=\Re\,e^{\Lambda}={\Re P_0 \over P}$.

\vspace*{12pt}
\noindent
{\tenbf Proposition 2:} If an AFRTEM spacetime possesses past marginally
trapped surface ${\cal T}_{u}$ on a null slice $u$, then the Bondi mass
${\cal M}_B$ at $u$ is positive and satisfies ${{\cal M}_B}^2 \geq Q_{0}^2$.

\vspace*{12pt}
\noindent
{\tenbf Proof:} We first show that the Bondi mass is non-negative.
(\ref{eq:mu}) gives
\begin{displaymath}
{2Me^{3\Lambda}} - {Q_{0}^2 e^{4\Lambda-\Phi}} = e^{\Phi}(1-\Delta_0 \Phi)
\end{displaymath}
The non-negativity of the Bondi Mass follows from this expression:
\begin{eqnarray*}
{\cal M}_B &=& {1\over 4\pi} \int_{S^2} {Me^{3\Lambda}}\,\omega_0  \\
&\geq &  -{1\over 2}\int_{S^2} e^{\Phi} \Delta_0 \Phi\,\omega_0    \\
&\geq &  -{1\over 2}\int_{S^2} [\Delta_0 (e^{\Phi})
         - 2e^{\Phi}(P_0 \Phi,_\zeta)(P_0 \Phi,_{\bar\zeta})]\, \omega_0  \\
&\geq & \int_{S^2} e^{\Phi}
                     (P_0 \Phi,_\zeta)(P_0 \Phi,_{\bar\zeta})\, \omega_0  \\
\Rightarrow {\cal M}_B &\geq & 0
\end{eqnarray*}

\noindent Since the electric charge $Q_0$ is assumed to be nonvanishing,
we can show by using the Schwarz inequality and
$\int_{S^2} e^{2\Lambda} \omega_0 = 4\pi$ that
\begin{eqnarray*}
16\pi^2 {{\cal M}_B}^2
&=& \left(\int_{S^2} Me^{3\Lambda}\,\omega_0 \right)^2    \\
&=& {1\over 4}\left(\int_{S^2} {Q_{0}^2}e^{4\Lambda-\Phi}\,\omega_0
  - \int_{S^2} e^{\Phi}(1-\Delta_0 \Phi)\,\omega_0 \right)^2   \\
& & \mbox{} + \left(\int_{S^2} {Q_{0}^2}e^{4\Lambda-\Phi}\,\omega_0\,\right)
              \left(\int_{S^2} e^{\Phi}(1-\Delta_0 \Phi\,)\,\omega_0\,\right)
\\
&\geq & \left(\int_{S^2} {Q_{0}^2}e^{4\Lambda-\Phi}\,\omega_0\,\right)
        \left(\int_{S^2} e^{\Phi}\,\omega_0 +
              \int_{S^2} e^{\Phi}
              (P_0 \Phi,_\zeta)(P_0 \Phi,_{\bar\zeta})\,\omega_0\,\right) \\
&\geq &
\left[\left(\int_{S^2}{Q_{0}^2}e^{4\Lambda-\Phi}\,\omega_0\,\right)^{1/2}
        \left(\int_{S^2} e^{\Phi}\,\omega_0\,\right)^{1/2} \right]^2  \\
&\geq & \left(\int_{S^2} Q_{0} e^{2\Lambda}\,\omega_0\, \right)^2     \\
&=& 16\pi^2 Q_{0}^2     \\
\Rightarrow {{\cal M}_B}^2 &\geq & Q_{0}^2 > 0 \hbox{\qquad\quad\qed\,.}
\end{eqnarray*}

\vspace*{12pt}
\noindent Proposition 2 is a special case of the
positive mass result of Gibbons et al.\/.$^{16}$

We can define the surface area of the marginally
trapped surface ${\cal T}_u$ as

\begin{equation}
{\cal A}_{\cal T} = \int_{S_{u,1}}\Re^2\,\omega_1
                  = \int_{S^2}  e^{2\Phi}\,\omega_0     \label{eq:at}
\end{equation}

\vspace*{12pt}
\noindent
{\tenbf Proposition 3:} If an AFRTEM spacetime possesses an outer past
apparent horizon ${\cal H}$, then the surface area ${\cal A}_{\cal T}$
of the outer marginally trapped surface ${\cal T}_{u}$ is a monotonically
decreasing function of the retarded time $u$.

\vspace*{12pt}
\noindent
{\tenbf Proof:} Apply (\ref{eq:mu}) to the inequality (\ref{eq:inequality}).
We get
\begin{equation}
(e^{2\Phi}),_u  + \Delta_0(e^{\Phi-\Lambda}) =
-e^{\Phi+\Lambda\/}({\widetilde{N_a}}{\widetilde{N^a}}) \leq 0
\label{eq:area1'}
\end{equation}

\noindent Since $\widetilde{N^a}$ is causal, by Stokes' Theorem

\begin{equation}
{d\over du} {\cal A}_{\cal T} = {d\over du} \int_{S^2} {e^{2\Phi}\,\omega_0}
=-\int_{S^2} e^{\Phi+\Lambda\/} ({\widetilde N}_a{\widetilde N}^a)\,\omega_0
\leq 0                             \label{eq:area2}
\end{equation}

\noindent \qed\,.

\vspace*{12pt}
\noindent Proposition $3$ is simply the
``area decrease theorem" for apparent horizons.$^{22}$\, If a spacetime
possesses a horizon, then as it evolves
the area of its past apparent horizon will
decrease, just as its
future horizon will increase in area. We can now prove that
the sum of the square of the Bondi mass and the area of the
outer marginally trapped surface at a retarded time $u$ is
a Lyapunov functional.

\vspace*{12pt}
\noindent
{\tenbf Theorem:} ${\cal L}(M, \Lambda, \Phi)= 16 \pi
{{\cal M}_B}^2 + {\cal A}_{\cal T}$ is a Lyapunov functional for a regular
AFRTEM spacetime that possesses an outer past apparent horizon ${\cal H\/}$;
i.e. ${\cal L}$ satisfies the following conditions:

\noindent (a) ${\cal L}(M, \Lambda, \Phi) > 0$,

\noindent (b) ${d\over du}{\cal L}(M, \Lambda, \Phi) \leq 0$, and

\noindent (c) ${d\over du}{\cal L}(M_0, \Lambda_0, \Phi_0) = 0$ if and only if
$(M_0, \Lambda_0, \Phi_0)$  is an equilibrium solution of the system.

\vspace*{12pt}
\noindent
{\tenbf Proof:}\par
\noindent (a) Proposition $2$ gives ${{\cal M}_B}^2 > 0$, and by definition
${\cal A}_{\cal T} > 0$. Hence ${\cal L} > 0$. \par

\noindent (b) Proposition $1$ gives ${d\over du}{\cal M}_B \leq 0$,
Proposition $3$ gives ${d\over du}{\cal A}_{\cal T} \leq 0$, and
Proposition $2$ gives ${\cal M}_B > 0$. Hence

\begin{displaymath}
{d\over du}{\cal L} = 32 \pi {\cal M}_B {d\over du} {\cal M}_B +
{d\over du}{\cal A}_{\cal T} \leq 0
\end{displaymath}

\noindent (c) (i) ($\Rightarrow$) \par
\quad ${d\over du}{\cal M}_B = 0 \Rightarrow  M,_\zeta =0$ and
$[P_{0}^2 (e^{-\Lambda}),_\zeta],_\zeta = 0 \,
$ (from (\ref{eq:dbm}) in Proposition $1$),  \par
\quad ${d\over du}{\cal A}_{\cal T} = 0 \Rightarrow \widetilde{N}_a
\widetilde{N}^a = 0$ \quad (from (\ref{eq:area2}) in Proposition 3)  \par
\qquad \qquad $\Rightarrow [P^2 (\Re^{-1}),_\zeta],_\zeta = 0\/$, and
${M,_\zeta} = {2Q_{0}^2 (\Re^{-1}),_\zeta}$\quad (from (\ref{eq:h3}))\par
\ So, ${d\over du}{\cal L} = 0 \Rightarrow {d\over du}{\cal M}_B = 0\/$, and
${d\over du}{\cal A}_{\cal T}=0 $\par
\qquad\qquad $\Rightarrow M,_\zeta = \Re,_\zeta
 = [P_{0}^2 (e^{-\Lambda}),_\zeta],_\zeta = 0 $\par
\ Now, $M,_\zeta = 0 \Rightarrow \Delta_0 M = 0$ , and \par
\qquad\qquad $[P_{0}^2 (e^{-\Lambda}),_\zeta],_\zeta = 0 \Rightarrow
\Delta_0 K = 0$ (from (\ref{eq:lgc}) in Proposition $1$)
\par
\qquad\quad$\Rightarrow M,_u = \Lambda,_u = 0$ (from (\ref{eq:rt1'}) and
(\ref{eq:rt2'})),\par
\qquad\quad$\Rightarrow M = M_0 = \hbox{constant,\ \ and\ \ }\Lambda =
\Lambda_0 = \ln P_0 - \ln (A+{\bar B}\zeta+B{\bar\zeta}+C \zeta
\bar\zeta)$.\par
\ Also, $ \Re,_\zeta = 0 \Rightarrow (e^{\Phi-\Lambda}),_\zeta = 0
\Rightarrow \Phi,_u = 0$ (from (\ref{eq:area1'}) in Proposition $3$).\par
Thus equilibrium is reached when ${\cal L}$ is stationary.
The equilibrium value of $\Re$ can be obtained from (\ref{eq:mu})
and (\ref{eq:h3}); i.e.
$\Re_{0}= M_0 + \sqrt{M_{0}^2-Q_{0}^2}$
\par
 (ii) ($\Leftarrow$)\par
\ $\Lambda = \Lambda_0, \quad M = M_0,
\quad \Re = \Re_0$\par
$\quad \Rightarrow {d\over du}{\cal M}_B = 0$ (from (\ref{eq:dbm})),
\quad and $\quad{d\over du}{\cal A}_{\cal T} = 0$ (from (\ref{eq:area1'})
and (\ref{eq:area2}))\par
$\quad \Rightarrow {d\over du}{\cal L} = 0$ \hbox{\qquad\quad\qed\,.}

\section{Conclusion}
\noindent
Rendall $^{23}$ pointed out, in relation to the AFRT evolution,
that the presence of a Lyapunov functional does
not guarantee the existence of a solution, as the phase space
is infinite dimensional.
Furthermore, due to the non-uniqueness of the equilibrium solution,
arising from conformal motions of the sphere,
the existence of a Lyapunov functional was not sufficient to
guarantee convergence to the equilibrium. An assumption of
antipodal symmetry would restrict the class of solutions such that there is
a unique equilibrium.$^{6}$\,
It has subsequently been suggested, in the vacuum case, that this extra freedom
can be factored out, effectively fixing a gauge condition.$^{9}$\,  It has
been shown that the conformal motions that preserve the equilibrium solution
are equivalent
to transformations at ${\cal I}^+$ between different accelerated frames.  We
will discuss
this point further in a future paper.
\par
We believe that the existence of a Lyapunov functional is very suggestive,
and that the AFRTEM system, in the case where a past apparent horizon exists,
evolves stably to the Reissner Nordstr\"om equilibrium.
(\ref{eq:h3}) implies that $\int_{S^2} (M-{Q_{0}^2 \over \Re})\,\omega_0
\geq 0$. The condition $M-{Q_{0}^2 \over \Re} \geq 0$, that is
${\widetilde{\Psi}}_2 \leq 0$, if it holds pointwise on
${\cal H}$, would be sufficient to rule out the exponentially
growing modes in (\ref{eq:eigen}). Note that
near equilibrium $M \geq {Q_{0}^2 \over \Re}$,
with equality holding in the extreme
Reissner Nordstr\"om geometry.
This is in contrast to the instability suggested by the linear analysis
(see Section 2 above).  We would infer from this
that the subclass of AFRTEM spacetimes that possess an outer past apparent
horizon are those that correspond to the stable submanifold of the linearized
system.  The requirement for the existence of an apparent horizon amounts to
the
imposition of a global condition selecting the physically meaningful solutions
of the evolution equations.  This would resolve the apparent contradiction
between the instability
of the linearized AFRTEM system and the stability of the perturbed
Reissner Nordstr\"om black hole, since only that subclass of the
AFRTEM spacetimes containing apparent horizons correspond to
perturbed black hole spacetimes.\par

\par
\nonumsection{Acknowledgements}
\noindent
We would like to thank Armen Khocharyan for helpful discussions
regarding the mathematical properties of Lyapunov functionals.

\nonumsection{References}

\end{document}

\bye